\documentclass[prl,aps,twocolumn,showpacs,floatfix,longbibliography]{revtex4-2}

\usepackage{microtype}
\usepackage[english]{babel}

\usepackage{mathrsfs}
\usepackage{rotating}
\usepackage{color}
\usepackage{graphicx}
\usepackage{dcolumn}
\usepackage{bm}
\usepackage{hyperref}
\usepackage{hypernat}
\usepackage[latin1]{inputenc}
\usepackage{pstricks}
\usepackage{amsmath,amssymb}
\usepackage{epstopdf}
\DeclareGraphicsExtensions{.pdf,.ps}

\usepackage{graphicx}
\usepackage{dcolumn}
\usepackage{bm}
\usepackage{ulem}

\begin{document}

\title{Nondipole Coulomb sub-barrier ionization dynamics and photon momentum sharing}

\author{Pei-Lun He}\email{peilun@mpi-hd.mpg.de}
\author{Michael Klaiber}
\author{Karen Z. Hatsagortsyan}
\email{k.hatsagortsyan@mpi-hd.mpg.de}
\author{Christoph H. Keitel}
\affiliation{Max-Planck-Institut f\"ur Kernphysik, Saupfercheckweg 1, 69117 Heidelberg, Germany}

\date{\today}

\begin{abstract}

The nondipole under-the-barrier dynamics of the electron during strong-field tunneling ionization is investigated, examining the role of the Coulomb field of the atomic core. The common analysis in the strong field approximation is consequently generalised to include the leading light-front non-dipole Coulomb corrections and
demonstrates the counter-intuitive impact of the sub-barrier Coulomb field. Despite its attractive nature, the sub-barrier Coulomb field increases the photoelectron nondipole momentum shift along the laser propagation direction, involving a strong dependence on the laser field. The scaling of the effect with respect to the principal quantum number and angular momentum of the bound state is found. We demonstrate that the signature of Coulomb induced sub-barrier effects can be identified in the asymptotic photoelectron momentum distribution via a comparative study of the field-dependent longitudinal momentum shift for different atomic species with state-of-the-art experimental techniques of mid-infrared lasers.

\end{abstract}

\maketitle

Exchanges of photon's energy, momentum, and angular momentum with charged particles are fundamental building blocks of light-matter interactions. In the presence of an intense laser field, photon energy absorption can trigger nonlinear processes like high-order harmonic generation \cite{huillier1993high-order,corkum1993plasma}, above-threshold ionization \cite{agostini1979free-free}, and nonsequential double ionization \cite{huillier1983multiply}.
Meanwhile, the angular momentum transfer from the laser beam is vital for generating vortex light \cite{marrucci2006optical,Chen_2018}, polarized electrons \cite{barth2011nonadiabatic,hartung2016electron} and positrons \cite{li2020production}, and nuclei vortices \cite{chen2020spiral}. The absorption of the photon momentum has attracted considerable interest recently due to various breakthroughs in the experimental technique \cite{smeenk2011partitioning,ludwig2014breakdown,Maurer_2018,hartung2019magnetic,Willenberg_2019,Haram_2019,Grundmann_2020,
Hartung_2021}.
High precision measurements raised questions on how the absorbed photon momentum is partitioned between the photoelectron and the parent ion. The analysis relies on investigating the photoelectron momentum distribution (PMD) along the laser propagation direction. The coincidence measurement of the photoelectron and its parent ion provides a direct check of the momentum conservation law in the microscopic world \cite{grundmann2020observation}.

It follows from energy and momentum conservation that in the simple-man model of tunneling ionization \cite{corkum1993plasma}, when the electron appears in the continuum with a vanishing momentum  due to a circularly polarized laser pulse, the total absorbed photon momentum is shared between the ion and electron as $I_p/c$ and $U_p/c$, respectively \cite{smeenk2011partitioning,Cricchio_2015}, where $I_p$ is the ionization energy, $U_p$ the ponderomotive potential, and $c$  the speed of light. However, in deviation to the simple-man picture, there is a longitudinal momentum transfer to the electron during tunneling due to the laser Lorentz force $\delta p_z \sim (\kappa/c)B_0\tau \sim I_p/c$, where $\kappa=\sqrt{2I_p}$ is the atomic velocity,
$\tau=\gamma/\omega$ the tunneling formation time, $\gamma=\omega \kappa/E_0$ the Keldysh parameter, $E_0$, $B_0$, and $\omega$ are the laser electric, and magnetic fields, and its frequency, respectively.
Atomic units are used throughout. Then, the most probable electron longitudinal momentum at the tunnel exit becomes $I_p/(3c)$  \cite{klaiber2013under},\cite{hartung2019magnetic}.
This is the case for a short-range atomic potential, when the electron begins the
sub-barrier motion at the bound state with a longitudinal momentum $v_{zs}=-2I_p/(3c)$ \cite{klaiber2013under,yakaboylu2013relativistic}.
The Coulomb field of the atomic core  reshapes the tunneling barrier and, consequently, the  nondipole momentum shift due to the sub-barrier dynamics. The latter is also affected by the momentum distribution of the bound state, which contributes into the birth probability of the quantum orbit at $v_{zs}$.

The long-range character of the Coulomb field of the atomic core is known to play a key role in strong-field ionization. It enhances the tunneling probabilities \cite{perelomov1966ionization,Perelomov_1966,Ammosov_1986,delone1991energy}.
In the continuum, it is responsible for the low-energy \cite{blaga2009strong,quan2009classical,Liu_2012,Wolter_2015,Liu_2010,Yan_2010,Kastner_2012} and high-energy structures \cite{keil2016laser,he2018high} in the PMD and the interplay between nondipole and Coulomb effects \cite{Forre_2006,ludwig2014breakdown,Maurer_2018,Danek_2018b,Danek_2018,Willenberg_2019h}. The sub-barrier dynamics can induce specific effects in tunneling ionization \cite{barth2011nonadiabatic,hartung2016electron,klaiber2013under,Yan_2012,Klaiber_2018}. For instance, the sub-barrier nonadiabatic dynamics results in a transverse momentum shift and polarized photoelectrons \cite{barth2011nonadiabatic,hartung2016electron}. The sub-barrier Coulomb effect can induce a phase shift of the quantum orbit \cite{Yan_2012}, disturbing the holography pattern. However, there is yet no conclusion on how the under-the-barrier Coulomb action affects the nondipole momentum shift. While simulations via the time-dependent Schr\"odinger
equation (TDSE) \cite{chelkowski2014photon,chelkowski2015photon,Chelkowski_2017} and time-dependent Dirac equation  \cite{Haram_2019}  give accurately the final longitudinal momentum shift, they do not provide  intuitive insight on the underlying  sub-barrier dynamics.

In this Letter, we investigate the electron's sub-barrier nondipole dynamics in strong-field tunneling ionization within an enhanced analytical description including the role of the Coulomb field of the atomic core.
The sub-barrier dynamics is described within the light-front \cite{dirac1949forms} nondipole Coulomb-corrected strong-field approximation (SFA), while the continuum one via the light-front classical trajectory Monte Carlo (CTMC) simulation, disentangling the effect of the sub-barrier Coulomb action from that in the continuum. We found that the sub-barrier Coulomb effect induces an increase of the photoelectron nondipole longitudinal momentum shift,
and discuss its counterintuitive physical origin. The dependence of the Coulomb effect on the angular momentum of the bound state is analyzed. Finally, we discuss how the Coulomb effects in the sub-barrier nondipole momentum shift can be observed experimentally using mid-infrared laser fields.

Our description of the tunneling process is based on TDSE  in the G\"oppert-Mayer gauge \cite{Reiss_1992}
\begin{equation}
i \frac{\partial}{\partial t} \psi(\mathbf{x}, t)=
\left[\frac{1}{2}\left(\textbf{p}+\mathbf{A} (\mathbf{x}, \eta)\right)^2+V(\textbf{x})-\phi(\mathbf{x}, \eta) \right]\psi(\mathbf{x}, t),
\label{sch}
\end{equation}
where $V(\textbf{x})=-\frac{Z}{|\mathbf{x}|}$ is the Coulomb potential of the atomic core, $A^\mu=(\phi,\mathbf{A})$ the laser four-vector potential in the G\"oppert-Mayer gauge, with $\phi=-\mathbf{x}\cdot \mathbf{E}(\eta)$ and $\mathbf{A} (\mathbf{x}, \eta)=\hat{\mathbf{z}}\phi/c$, $\mathbf{E}(\eta)=E_0\left[\hat{\textbf{x}}\cos(\omega \eta)-\hat{\textbf{y}}\sin(\omega \eta)\right]\exp\left(-2\ln2 \frac{\eta^2}{L^2}\right)$ the laser field of  circular polarization with $L=4~T$ and $T$ is the optical cycle, and with the light-front coordinate $\eta=t-z/c$.
The Hamiltonian in Eq.~(\ref{sch}) is accurate up to the $1/c$ order, suitable in the considered parameter regime.
 The deviation from the fully relativistic theory \cite{popov1997imaginary,mur1998relativistic,milosevic2002semiclassical,klaiber2013above} is given by $I_p/c^2$ terms for the sub-barrier and $U_p/c^2$ terms for the continuum dynamics. We apply light-front nondipole Coulomb-corrected SFA \cite{supp}, which corresponds to the nondipole expansion of the relativistic theory \cite{klaiber2013above}.
The nondipole transition amplitude in the G\"oppert-Mayer gauge reads  \cite{he2017strong}:
\begin{equation}
\begin{aligned}
M_\mathbf{p}&=-i\int d\eta \frac{d^3\mathbf{x}}{(2\pi)^{3/2}}
%(1-\widetilde{p}_z/c)
 e^{-i\int_{\eta}^{\infty}d{\eta}'\frac{\frac{1}{2}\widetilde{\mathbf{p}}({\eta}')^2+I_p}{1-\frac{\widetilde{p}_z}{c}}}  {\cal F}^*_C\left(\eta,\mathbf{x}\right) \\
& \times (1-\widetilde{p}_z/c)  e^{-i \widetilde{\mathbf{p}}(\eta) \cdot \mathbf{x} } \mathbf{x}\cdot \mathbf{E}(\eta) \psi_{\kappa l m}(\mathbf{x}) ,
%{\cal F}^*_C\left(\eta,\mathbf{x}\right),
\end{aligned}
\label{transition}
\end{equation}
where  $\widetilde{\mathbf{p}}(\eta)=\left(\mathbf{p}_\perp +\mathbf{A}_\perp({\eta}),p_--\frac{I_p}{c}\right)$, with $ p_-=p_z-\frac{\mathbf{p}^2}{2c}$ is the light-front momentum, $\mathbf{A}_\perp({\eta})\equiv -\int_\infty^\eta d\eta' \mathbf{E}(\eta')$, and $\psi_{\kappa l m}(\mathbf{x})$ is the bound state wave function. Note that $p_z/c\approx p_-/c \approx \widetilde{p}_z/c$ up to the $1/c$ order.
The use of the G\"oppert-Mayer gauge in the employed SFA is justified as it provides the accurate Coulomb corrected Perelomov-Popov-Terent'ev (PPT) ionization rate \cite{perelomov1966ionization,Perelomov_1966,Ammosov_1986}, which is well confirmed experimentally.
 The influence of the Coulomb field on the sub-barrier continuum electron is described by the Coulomb correction (CC) factor:
\begin{equation}
\label{QQQ}
 {\cal F}_{C} \left(\eta,\mathbf{x}\right) =
\exp\left\{i \int_{\eta }^{\infty}d{\eta}'  V\left(\mathbf{x}(\eta,{\eta}')\right)/\left(1- \frac{p_-}{c}\right)\right \}.
\end{equation}
With saddle-point integration of Eq.~(\ref{transition}), the differential ionization rate in the adiabatic regime is obtained \cite{supp}:
\begin{equation}
\begin{aligned}
\Gamma^{\left(n,l,m\right)} &\left(\eta,v_\perp,p_-\right)
=  \frac{\kappa}{\widetilde{\kappa}} \frac{2^{2\nu+2} C^2_{\kappa l}\kappa^{6\nu} }{(1-\frac{p_-}{c})^{2\nu}E(\eta)^{2\nu}} \\
&\times \left |Y_{lm}\left(\widetilde{\mathbf{p}}(\eta_s)\right)\right |^2 \exp\left[-\frac{2}{3}\frac{\widetilde{\kappa}^3}{E(\eta)(1-\frac{p_-}{c})} \right].
\end{aligned}\label{rate}
\end{equation}
where $\left(n,l,m\right)$ are the atomic quantum numbers (the angular momentum projection is determined with respect to the laser propagation axis), $E(\eta)=\left |\mathbf{E}(\eta)\right |$, $\widetilde{\mathbf{p}}(\eta_s)$ is the saddle-point momentum, $\nu=Z/\kappa$ the effective principal quantum number, $v_\perp$ the initial velocity component in the polarization plane, $C_{\kappa l}^{2}=\frac{2^{2 \nu-2}}{\nu \left(\nu+l\right) !\left(\nu-l-1\right) !}$, and $\widetilde{\kappa} = \sqrt{\kappa^2+\widetilde{p}_z^2+v^2_\perp}$.
\begin{figure}[t]
\begin{center}
\includegraphics[width=0.50\textwidth]{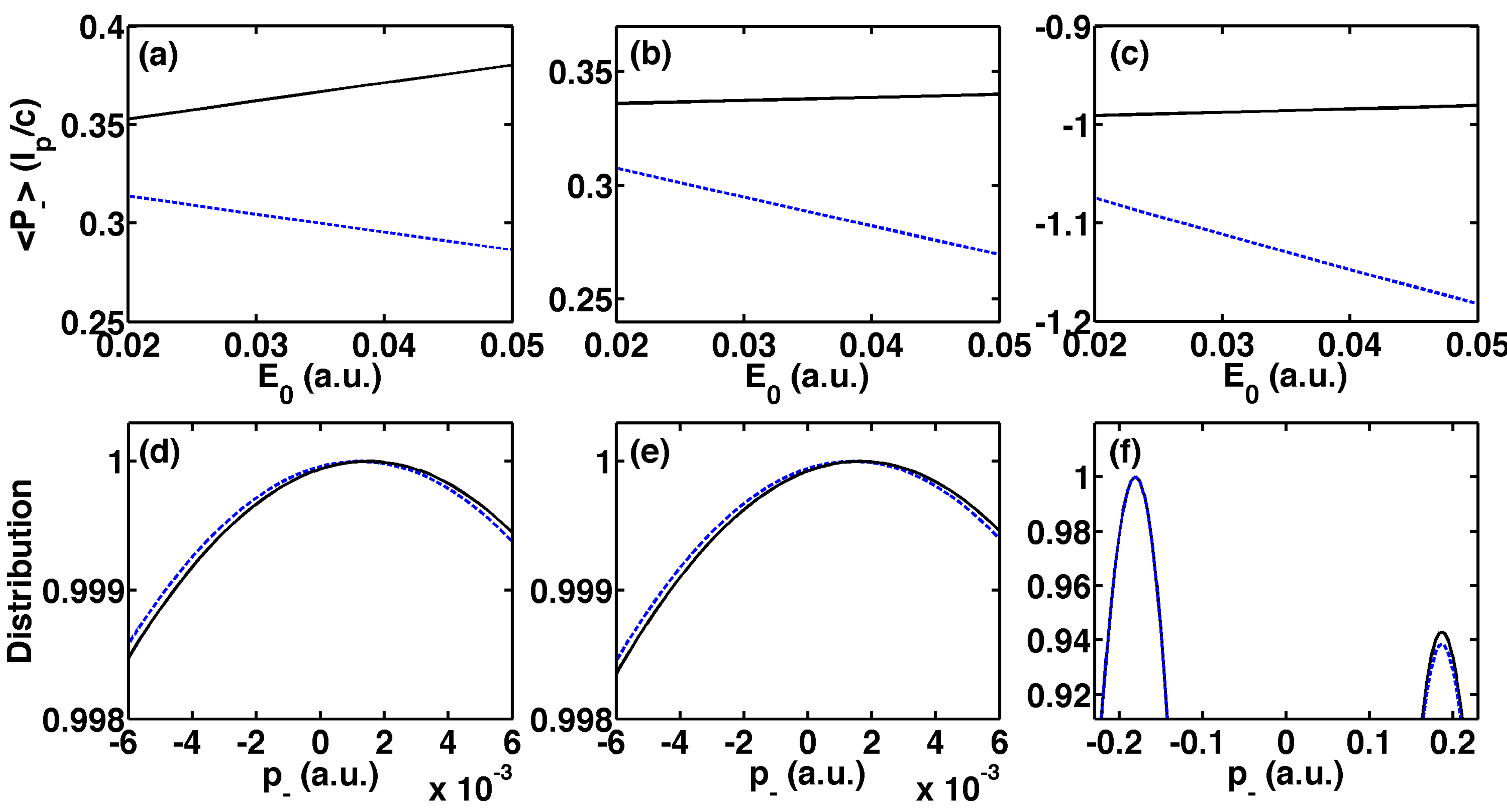}
  \caption{The sub-barrier Coulomb effect, neglecting the continuum influence, in the quasistatic regime:
(a-c) The expectation value $\left \langle  p_- \right \rangle $ vs $E_0$; (d-f) PMDs at the laser intensity $I=10^{14}$~W/cm$^2$. The initial states are (a,d) hydrogen atom with quantum numbers $(1,0,0)$, and argon atom
(b,e)  with  $(3,1,1)$, and (c,f) with  $(3,1,0)$. The solid-black line includes the sub-barrier Coulomb effect, while the dashed-blue line is the case of a short-range potential.
}
 \label{fig1}
\end{center}
\end{figure}

To implement the light-front CTMC simulations, we  sample over all possible ionization instants, tunneling exits, and initial momenta as initial conditions for the classic equation of motion
\begin{equation}
\begin{aligned}
\frac{dQ}{d\eta}=\{Q,H\}, ~~ \frac{dP}{d\eta}=\{P,H\},
\end{aligned}
\end{equation}
where $\{A,B\}$ is the Poisson bracket, with the canonical coordinates $Q=\left(x,y,z\right)$ and $P=\left(p_x,p_y,p_-\right)$. Trajectories are weighted by $\Gamma^{\left(n,l,m\right)}$ of Eq. (\ref{rate}).
Accurate to the $1/c$ order, we have the classical light-front Hamiltonian
\begin{equation}
H=\frac{\left(\mathbf{p}_{\perp}+\mathbf{A}_{\perp} \right)^2+\left(p_-+V/c\right)^2}{2\left(1-\frac{p_-+V/c}{c} \right)}+V.
\end{equation}
The tunneling exit $\mathbf{x}^e=-\frac{I_p}{1- p_-/c}\frac{\mathbf{E}(\eta)}{E(\eta)^2}$ is obtained by taking the real part of the complex trajectory \cite{supp}.

Let us first discuss the sub-barrier Coulomb effect. Neglecting the Coulomb field effect in the continuum leads to the conservation of $P=\left(p_x,p_y,p_-\right)$. We plot the expectation value of $p_-$ in Fig.~\ref{fig1}.
From Eq.~(\ref{rate}),  we derive $\langle p_-\rangle$ for an $s$-state in the leading order correction of $E_0/E_a$ \cite{supp}:
\begin{equation}
\left \langle p_- \right \rangle=\frac{I_p}{c}\left[\frac{1}{3}+\left(2\nu-1\right)\frac{E_0}{E_a}\right], \label{meanshift}
\end{equation}
with the atomic field $E_a=\kappa^3$.  Note that the leading $I_p/(3c)$ term can be derived only employing the binding energy $I_p$ and classical Lorentz sub-barrier dynamics in imaginary-time without any extra information about the binding potential and any time or spatial dependencies of the laser field, see Sec.IB.2 in \cite{supp}. Apart from the constant $I_p/(3c)$ \cite{klaiber2013under,chelkowski2014photon,he2017strong,hartung2019magnetic}, the sub-barrier Coulomb effect yields an increase in the momentum shift, despite the attractive nature of the parent ion, and
induces a strong dependence on the laser field and the principal quantum number of the bound state [Fig.~\ref{fig1}].

The active electrons of most rare gas atoms are in the $p$-state. For instance, the ionization of the argon atom is dominated by signals from $(3,1,\pm 1)$ orbits, while the nondipole shift for $(3,1,0)$ is more distinctive:
\begin{equation}
\left \langle p_- \right \rangle=\frac{I_p}{c}\left[-1+\left(6\nu-5\right)\frac{E_0}{E_a}\right].
\end{equation}

\begin{figure}
\begin{center}
  \centering
\includegraphics[width=0.5\textwidth]{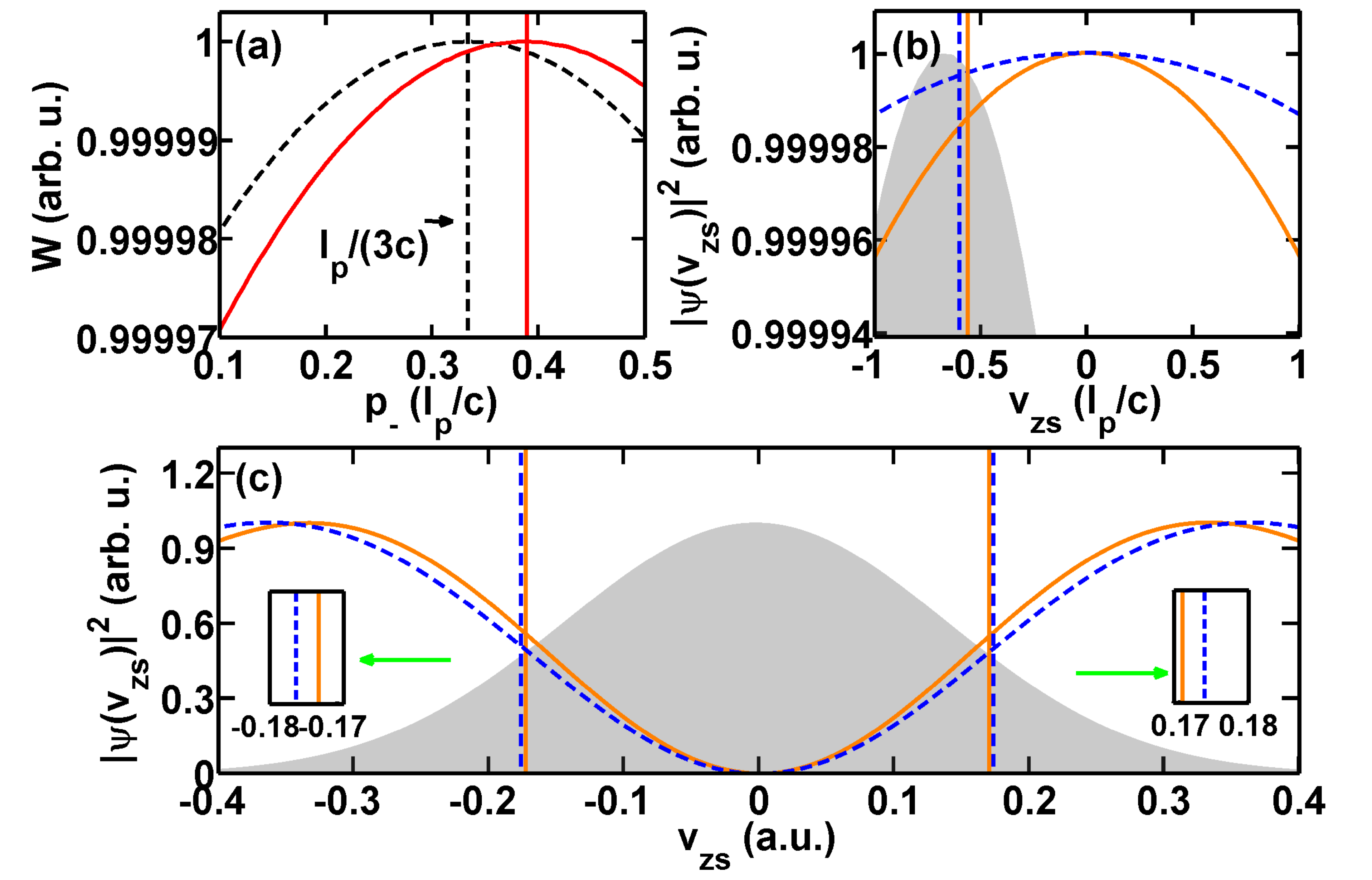}
   \caption{
   (a) The nondipole CC ionization window ${\cal W}(p_-)\equiv\left|{\cal F}_C\right|^2\exp\left[-\frac{2}{3}\frac{\widetilde{\kappa}^3}{E(\eta)(1-p_-/c)}\right]$: (solid-red) including the sub-barrier action, the peak is at $p^m_- > I_p/(3c)$  (solid-red gridline), (dashed-black) neglecting the sub-barrier Coulomb correction, $p^m_- \approx  I_p/(3c)$ (dashed-black gridline). The bound state wave functions at the starting point of the tunneling trajectory $x_s$: (b)  for $(1,0,0)$ state, (c) for $(3,1,0)$ state; the wave function for the zero-range potential (dashed-blue), and for the Coulomb case (solid-orange); the tunneling window ${\cal W}$ is shadowed;  $E_0=0.04$ a.u. The gridlines in (b-c) indicate the peaks of ${\cal W}(v_{zs})|\psi(x_s,y_s,v_{zs})|^2$ with (solid-orange), and without the sub-barrier CC (dashed-blue). }
    \label{fig2}
\end{center}
\end{figure}

The physical origin of the Coulomb field effects for the nondipole sub-barrier dynamics is encoded in the exponent and the prefactor of the ionization amplitude of Eq.~(\ref{transition}). The action contributed by the kinetic part, i.e. $\frac{\frac{1}{2}\widetilde{\mathbf{p}}({\eta}')^2+I_p}{1-\frac{\widetilde{p}_z}{c}}$ of Eq.~(\ref{transition}), gives the tunneling exponent in Eq.~(\ref{rate}). Due to the laser magnetic field, the tunneling electron obtains a momentum in the laser propagation $z$-direction, which decreases the longitudinal tunneling energy and the tunneling probability, the latter being maximal at $p^m_-=I_p/(3c)$ \cite{klaiber2013under}. We can estimate the contribution of the Coulomb action
by evaluating Eq.~(\ref{QQQ}) with quantum orbits starting at the saddle point of the SFA-matrix element $\mathbf{x}_s$ \cite{klaiber2017strong}.
The momentum shift due to the sub-barrier Coulomb action ${\cal F}_C$ is evaluated by the $\eta$-integration from the saddle-time $\eta_s$ to the exit time $\eta_r=\rm{Re} ~\eta_s$.
%calculating  ${\cal F}_C\equiv {\cal F}_{C}\left(\eta_s,\mathbf{x}_s\right)/{\cal F}_{C}\left(\eta_r,\mathbf{x}_s\right)$, with  the saddle-time $\eta_s$, the exit time $\eta_r=\rm{Re}\{\eta_s\}$, using the sub-barrier trajectory starting at the saddle point of the SFA-matrix element $\mathbf{x}_s$ \cite{klaiber2017strong}.
%When we include the full sub-barrier dynamics, motions along the tunneling channel ($x$-direction)  as well as along the laser propagation direction ($z$-direction), then the CC factor ${\cal F}_C$ does not change much the most probable momentum for the case of a short-range atomic potential $p^m_-=I_p/(3c)$ \footnote{However, the CC in $x$-direction  increases the nondipole momentum shift (this corrects the simple estimate in \cite{yakaboylu2013relativistic}), but the sub-barrier $z$-dynamics compensates that.}, see Fig. \ref{fig2}(a). Thus, the definition of the tunnel exit and the separation of $\eta$-integral in ${\cal F}_C$ into the sub-barrier and continuum parts is not relevant for emerging of the sub-barrier CC.
The sub-barrier Coulomb action ${\cal F}_C $ contributes to the momentum shift, yielding a peak 
%When we include the sub-barrier action, the peak is 
at $p^m_- > I_p/(3c)$, see Fig. \ref{fig2}(a).
An essential sub-barrier CC for the nondipole momentum shift stems from matching the bound state wave function in the Coulomb field to the tunneling wave packet.
%Matching the bound state wave function in the Coulomb field to the tunneling wave packet also contributes to the momentum shift.
The prefactor accounts for the probability of the quantum orbit, which originates at the bound state with the momentum $v_{zs}$. The width of the momentum space wave function $\psi_{\kappa lm}(x_s,y_s,v_{zs})$ (in the mixed representation, at the starting point of the ionization $x_s$) in the Coulomb field is narrower than that in the short-range potential, see Fig.~\ref{fig2}(b).
The most probable $v^m_{zs}$ is the peak of the function: $|\psi(x_s,y_s,v_{zs})|^2$ ${\cal W}(v_{zs}$), which is indicated by the gridlines in Fig.~\ref{fig2} (b,c). The solid-orange grid line in (b) includes the sub-barrier CC and is closer to the origin (larger $v_{zs}$), which yields an increase of the longitudinal momentum at the tunnel exit, as the total longitudinal momentum transfer from $\mathbf{x}_s$ to $\mathbf{x}^{e}$ is $I_p/c$.

\begin{figure}[b]
\begin{center}
 \centering
\includegraphics[width=0.5\textwidth]{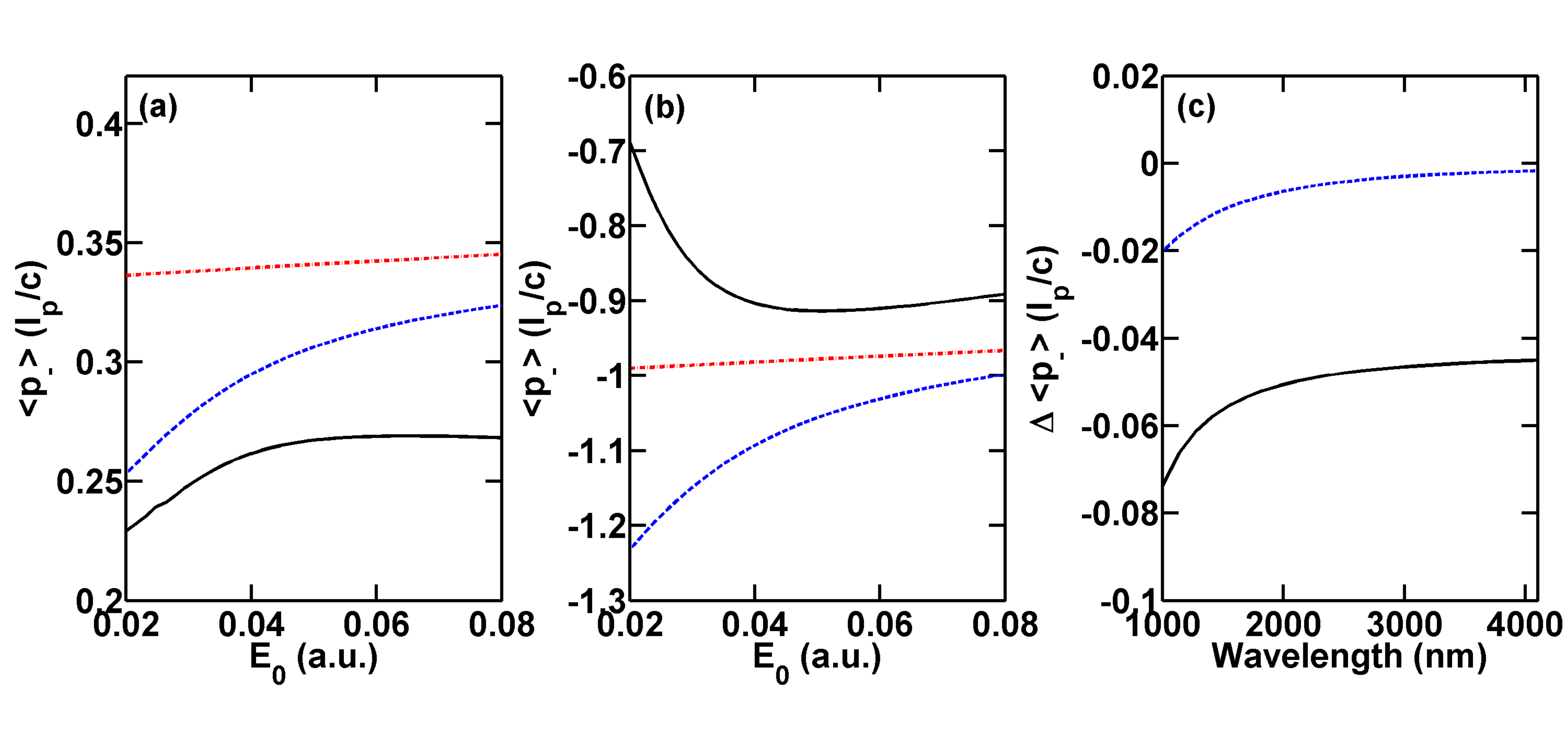}
  \caption{ Asymptotic light-front momentum $\left \langle p_- \right \rangle $ vs $E_0$
when the initial states are an argon atom with quantum numbers (a) $\left(3,1,1\right)$ and (b) $\left(3,1,0\right)$. The solid black lines include the continuum and the sub-barrier CC, the dashed-dotted red lines only the sub-barrier CC, and the dashed blue lines only the sub-barrier CC and nonadiabatic effects.
(c) The contributions $\Delta \left \langle p_- \right \rangle $ of the nonadiabatic effect (dashed-blue) and the continuum CC (solid-black) to $\left \langle p_- \right \rangle $ vs the laser wavelength, for the laser intensity of $10^{14}$ W/cm$^2$.}
 \label{fig3}
\end{center}
\end{figure}

The momentum shift and the corresponding CC depend essentially on the angular momentum of the bound state.
We compare in Fig.~\ref{fig1} the expectation value of $\left \langle p_- \right \rangle$ with and without CC for
quantum numbers $ (1,0,0 )$ (hydrogen) and $ (3,1,0 ), \,(3,1,1 )$ (argon).
There is a significant difference between $s$- and $p$-states.
 While the PMD of $(1,0,0 )$ and $(3,1,\pm 1 )$ has only one peak, it has two peaks around $\pm \sqrt{E_0/\kappa}$ for $(3,1,0)$, which originate from the momentum distribution of the initial bound state, see Fig.~\ref{fig2}(c).
In the dipole theory, the two peaks are equally probable and $\langle p_- \rangle=0$.
The center of the window ${\cal W}$ is at $v_{zs} =-2I_p/(3c)$ in the nondipole theory. The asymmetry of the window with respect to the origin results in suppression of the probability of the positive peak with respect to the negative one, yielding a negative longitudinal momentum $\langle p_- \rangle <0$.
After accounting for the atomic Coulomb field, the peaks of the wave function become closer, which results in an increase of the PMD peaks.
The sub-barrier CC for the right peak is larger [Fig.~\ref{fig1}(f)], which leads to a decrease of $\langle p_- \rangle $ by absolute value. Hence for the state $(3,1,0)$, the exponent of Eq.~(\ref{rate}) prefers the peak $p_-\approx -\sqrt{E_0/\kappa}$ due to the nondipole momentum shift $I_p/c$, while the prefactor prefers the peak $p_-\approx \sqrt{E_0/\kappa}$ due to the sub-barrier Coulomb field effect.

\begin{figure}[t]
\begin{center}
\includegraphics[width=0.5\textwidth]{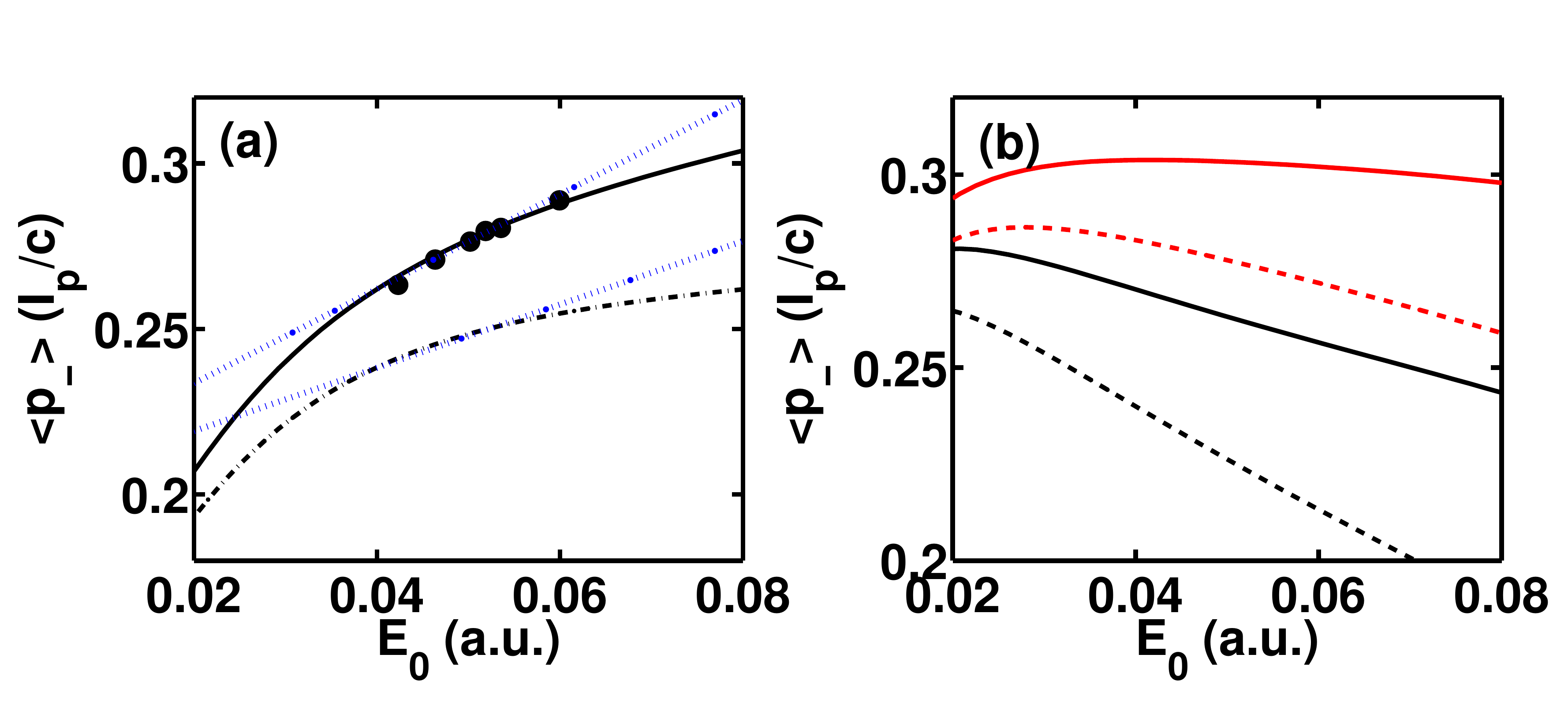}
 \caption{  (a) $\left \langle p_- \right \rangle $ vs $E_0$ for the ground state hydrogen atom when the laser wavelength is 800 nm.
%The solid lines include the full CC, while the dashed lines only the continuum CC.
Circles are results of TDSE simulations. Dotted blue lines are linear fittings of the curves.
(b) $\left \langle p_- \right \rangle $ vs $E_0$ for Xe and Xe$^+$ when the laser wavelength is 3000 nm. The solid lines include the full CC, while the dashed lines do not include the sub-barrier CC.
The black lines are for Xe, and the red lines are for Xe$^+$.}
\label{fig4}
\end{center}
\end{figure}

Next let us discuss the role of the Coulomb field in the continuum and the nonadiabatic effect, see Fig.~\ref{fig3}. $P=\left(p_x,p_y,p_-\right)$ is no longer conserved  and the momentum is reduced by absolute value as a result of the Coulomb momentum transfer mostly at the tunnel exit.
For the states $(3,1,\pm 1)$ (argon) the continuum CC decreases the momentum shift, however, approximately maintaining the slope of the field dependence \cite{supp}.
For a smaller electric field strength, the tunneling exit is farther away from the atomic core, while the photoelectron can stay a longer time around the nuclei. As a consequence, $\left \langle p_- \right \rangle $ decreases in the absolute value, while the slope of the curve remains less affected.
%The continuum CC is smaller in the case of $(3,1,\pm 1)$ state than for $(1,0,0)$. This is because for $(3,1,\pm 1)$ the angular momentum brings prefactor $\left(v_\perp\mp \kappa \right)^2$, and the most probable initial momentum at the tunnel exit is $v_{\perp}=\mp \frac{E}{\kappa^2}$ for $(3,1,\pm 1)$, i.e., the electron leaves the parent ion rapidly, less affected by the Coulomb field. If the sub-barrier CC is neglected, the field dependence of the asymptotic longitudinal momentum shift is qualitatively different from the full CC case. In particular, it saturates at high field strengths in the case of $(1,0,0)$, or changes slowly in the case of $(3,1,\pm 1)$ and $(3,1,0)$ states.
For the state $(3,1,0)$, the continuum CC reduces the momentum shift for the peak at $p_-\approx \sqrt{E_0/\kappa}$ while increasing it for the peak at $p_-\approx -\sqrt{E_0/\kappa}$. The net result increases the total momentum shift.
The qualitative behavior of the nonadiabatic corrections is similar 
%adiabatic and the nonadiabatic theory is the same 
(those without sub-barrier CC were considered in \cite{Ni_2020}).
While the nonadiabatic effect decreases the absolute value of $\left \langle  p_- \right \rangle $, it increases the slope of the field dependence. 
%curve. To quantitatively calibrate the contribution of nonadiabatic effects and the continuum CC, we compare the difference of the momentum shift $\Delta \left \langle p_- \right \rangle$ obtained from different theory in Fig. \ref{fig3}(c). A longer laser wavelength suppresses both the continuum CC and the nonadiabatic effect.
The contributions of the nonadiabatic and the continuum CC effects are calibrated via the difference of the momentum shift $\Delta \left \langle p_- \right \rangle$, calculated with and without these effects, see Fig.~\ref{fig3}(c). Both the continuum CC and the nonadiabatic effects are suppressed at longer  laser wavelengths.

It is possible to detect the sub-barrier CC signal with the current experimental capabilities similar to Refs.~\cite{hartung2019magnetic,Willenberg_2019}.
%However, one will need to collect more data with different atoms and with a large range of accessible laser intensities.
The momentum shift as well as the slopes of the field dependence are different after including the sub-barrier effect.
In Fig.~\ref{fig4}(a), we sample the  laser intensity range $10^{14}-2 \times 10^{14}$ W/cm$^2$ for the linear fitting.
We have carried out TDSE simulations based on the split-operator method, from which we obtain the scaling of the momentum shift as $\left \langle p_- \right \rangle \sim 1.44 \pm 0.22 ~(E_0/E_a)(I_p/c)$.
After including both the nonadiabatic effect and Coulomb corrections, the light-front CTMC simulations yield $\left \langle  p_- \right \rangle \sim 1.42 \pm 0.02 ~(E_0/E_a)(I_p/c)$.
Without the sub-barrier CC, the CTMC simulation gives only $\left \langle  p_- \right \rangle \sim 0.96 \pm 0.02 ~(E_0/E_a)(I_p/c)$.
Thus, the sub-barrier Coulomb dynamics induces a quite strong field dependence in the nondipole longitudinal momentum shift (a large slope vs the laser field) partly distinguishable in an experiment.

Comparing atoms with different effective principal quantum numbers $\nu$ provides another possibility to observe the sub-barrier CC, see Fig.~\ref{fig4}(b).
The qualitatively robust signature of the sub-barrier CC is the increase of $\left \langle p_- \right \rangle \sim 2\nu\frac{E_0}{E_a}\frac{I_p}{c}$  with a higher effective principal quantum number of the active electron.
For the ionization of Xe and Xe$^+$ with quantum numbers $(3,1,1)$, we have $\nu=1.06$ and $\nu=1.60$, respectively.
The sub-barrier CC signal is robust against the average over the magnetic quantum numbers.
Thus, the comparison of the longitudinal momentum shift and its dependence on the laser field for different atoms can unambiguously confirm the signature of the sub-barrier CC in an experiment.
The longer laser wavelengths are more beneficial for the observation of the considered effects, as the masking effect of nonadiabaticity can be avoided at $\gamma<1$, see Fig.~\ref{fig3}(c), which shows an increase of the continuum CC due to nonadiabaticity at shorter wavelengths.

Concluding, after introducing the light-front non-dipole Coulomb-corrected theory, we reveal a counterintuitive role of the Coulomb field of the atomic core in nondipole sub-barrier dynamics, which increases the photoelectron nondipole momentum shift despite its attractive nature, with consequences for the momentum partition of the absorbed laser photons. We demonstrate, disentangling the sub-barrier and continuum effects, that the Coulomb sub-barrier signatures  can be observed in the asymptotic momentum distribution with long-wavelength laser fields by measuring the characteristic slope of the field dependence of the average longitudinal momentum in different gas targets.

P.-L. H. acknowledges support via the prize of the Shanghai Jiao Tong University for outstanding graduate students. We thank Manfred Lein for a helpful comment regarding the effective principal quantum number.

\bibliography{references}

\end{document}